# A quantum mechanical scheme to reduce radiation damage in electron microscopy

Hiroshi Okamoto*, Tatiana Latychevskaia and Hans-Werner Fink

Institute of Physics, University of Zurich

Winterthurerstrasse 190, Zurich CH-8057, Switzerland

Author to whom correspondence should be addressed: okamoto@physik.uzh.ch

## Abstract

We show that radiation damage to unstained biological specimens is not an intractable problem in electron microscopy. When a structural hypothesis of a specimen is available, quantum mechanical principles allow us to verify the hypothesis with a very low electron dose. Realization of such a concept requires precise control of the electron wave front. Based on a diffractive electron optical implementation, we demonstrate the feasibility of this new method by both experimental and numerical investigations.

## Main text

It is well recognized that radiation damage severely restricts our ability to study unstained biological specimens in electron microscopy, thereby constituting a major obstacle in structural biology (1). The dilemma is that a low electron dose results in poor image quality, while a high dose leads to destruction of specimens. As such, it is widely believed that the problem is of a fundamental nature and that essentially no solution exists. The principal aim of our Letter is to show that this is not the case, and that one can circumvent the problem by taking advantage of prior knowledge to reduce the statistical noise associated with the detection of individual electrons. Such a reduction of electron dose is possible by measuring the scattered electron wave packets one by one in a more intelligent way than simple projection onto a screen. The reduction of electron dose for a given amount of retrieved structural information will ultimately translate to a better resolution. This concept is most drastically illustrated when the experimenter already has a correct hypothesis of the specimen structure, or more precisely, the presumed "image" including phase information that would be obtained if there were no radiation damage. In the present Letter we focus on this particular case. We emphasize that our method is not based on averaging; rather, it is applied to individual specimens.

Quantum mechanics states that a measurement of an observable produces a deterministic outcome when a particle is in an eigenstate of the observable. Repetition of the measurement on particles prepared in an identical way quickly yields high confidence about the quantum state due to the absence of noise. The same principle can be applied to the case of electron microscopy, if the state of an electron after the scattering by a specimen is known. One can then design a corresponding electron-optical element that transforms the scattered wave, i.e., the object wave, to a converging spherical wave to produce a spot on an imaging screen. Therefore it is possible to be highly confident about a given structural model with a single electron only, because the chance of detecting an electron at a particular point on the screen by coincidence is low. By repeating the measurement, confidence about the structural model will increase rapidly. This is in a sharp contrast to conventional electron microscopy, which typically takes $10^8$ electrons to generate an image with a fair signal to noise ratio.

Figure 1 shows how to transform the object wave $\psi_O$ from a specimen to a converging spherical wave $\psi_S$ by using a diffractive element (2). Coherence of the electron wave is essential here to interact with the whole specimen. We obtain $\psi_O$ by illuminating the specimen with a primary incident electron beam with a size just large enough to contain the specimen. The transmission coefficient distribution $d$ of the diffractive element has to be computed numerically in accordance with the structural hypothesis to be tested. Ideally, $d$ should provide both amplitude and phase modulation, satisfying $\psi_O \cdot d = \psi_S$ at the plane where the diffractive element is placed. Thus, by illuminating the diffractive element with the object wave $\psi_O$, the converging spherical wave emerges from the diffractive element. However, d has to be real and positive because a diffractive element is an amplitude object. Our approach is to take the real part and then add a constant so that

$$d = \text{Re}\left(\frac{\psi_S}{\psi_O}\right) + \text{const} \qquad (1)$$

The constant is chosen to ensure that $d$ is always positive. To avoid division by zero, $d$ was set to zero where the object wave intensity $|\psi_O|^2$ is smaller than a certain threshold. The emerging wave $\psi_O \cdot d$ is the sum of $\psi_S$ and other terms. The structural complexity of the diffractive element depends on the number of pixels in the image to be verified. The required number of "pixels" for the diffractive element to be fabricated is of the same order as the image pixel number.

The use of coherent electrons for structural determination *per se* is not new. Examples include the recent proposal of electron diffraction microscopy (3).

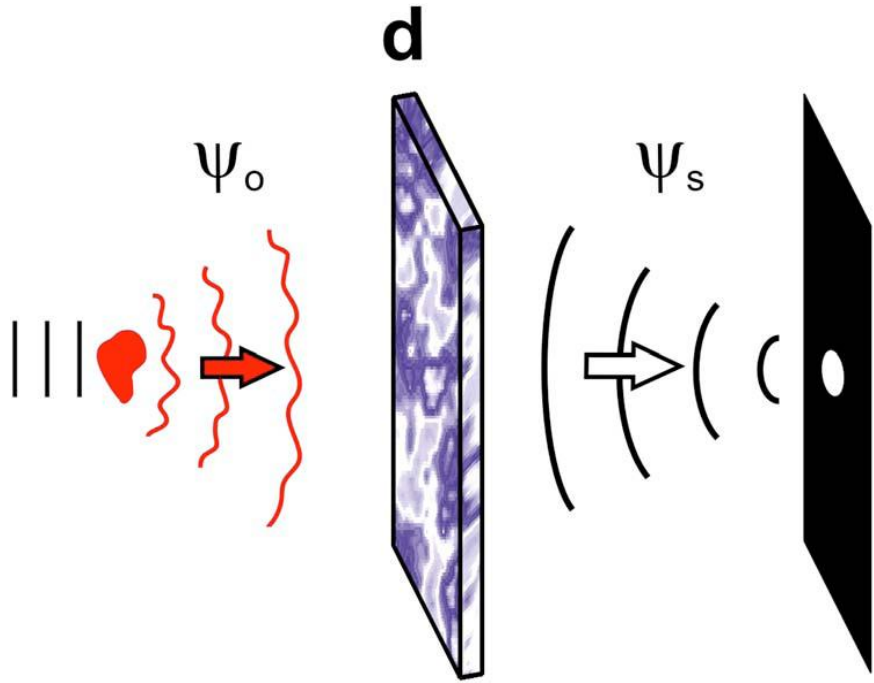

Figure 1. The scheme to transform the object wave $\psi_O$ from a specimen to a converging spherical wave $\psi_S$ by use of the diffractive element d. The structure of the diffractive element is computed numerically in accordance with Eq. (1).

In general, we do not know a priori the exact place and orientation of a specimen, say, a macromolecule. While small changes to the specimen position translate to the lateral displacements of the spot on the screen, the uncertainty of the specimen orientation poses a serious problem. Methods such as trapping of the molecules in a well-defined manner may have to be employed. Alternatively, it may be argued that our method is able to separate the "lucky" cases from the others, while conventional imaging methods damage each specimen by an average amount with certainty.

To realize the above scheme, we need freestanding diffractive elements that can be designed and fabricated. Indeed, it has been demonstrated that fabrication of such an element is possible (4). Here we show that diffractive elements can be fabricated straightforwardly even for low energy electrons, where a higher contrast of biological specimens is expected. This situation is due to the recent wide availability of commercial focused ion beam (FIB) apparatus. Here we demonstrate electron diffraction by gratings, as well as rudimentary lens action by a diffractive element. Figure 2(a) shows the fabricated slits of the gratings as imaged by low energy electron point source (LEEPS) microscopy (5). They were fabricated by a FIB apparatus (Orsey Physics), using a 30 keV, 2 pA Ga ion beam. Holes were milled in carbon films. The slit-to-slit distance was 100 nm. The array of four slits was approximated by a 4 × 4 array of approximately 20 nm diameter holes with anisotropic hole-to-hole spacing, since it provided better mechanical strength. The experimental setup is illustrated in Fig. 2(b). Since the characteristic feature size of the fabricated structures exceeds typical low energy electron wavelengths by a factor of several hundreds, we must stay in the paraxial optical regime. The high spatial coherence of the electron beam from the sharp field emitter guarantees the electron wave over the whole diffraction grating to be coherent. The relatively large distance of 360 μm between the electron source and the diffraction element ensured that, despite divergence from the point source, the angular spread of the incoming beam was smaller than the expected diffraction angle. The distance between the electron source and the screen was 100 mm, implying that our concept can be implemented without resorting to large instrumentation. Figure 2(c) shows a resultant image of the diffraction experiment performed with 149 eV electrons. The first order diffraction spots are clearly visible at an angle of 1.9 mrad from the zeroth order spots. This is close to the expected angle of 1.7 mrad derived from the measurement of the slit geometry, taking the electron beam divergence and the finite slit width into account. Furthermore, this angle changed consistently with the electron wavelength, which was varied from 0.073 to 0.13 nm, thus confirming that these spots are indeed due to diffraction. Figure 2(d) shows a rudimentary lens action, where the two diffraction spots merged into one when the electron energy was lowered to 90 eV.

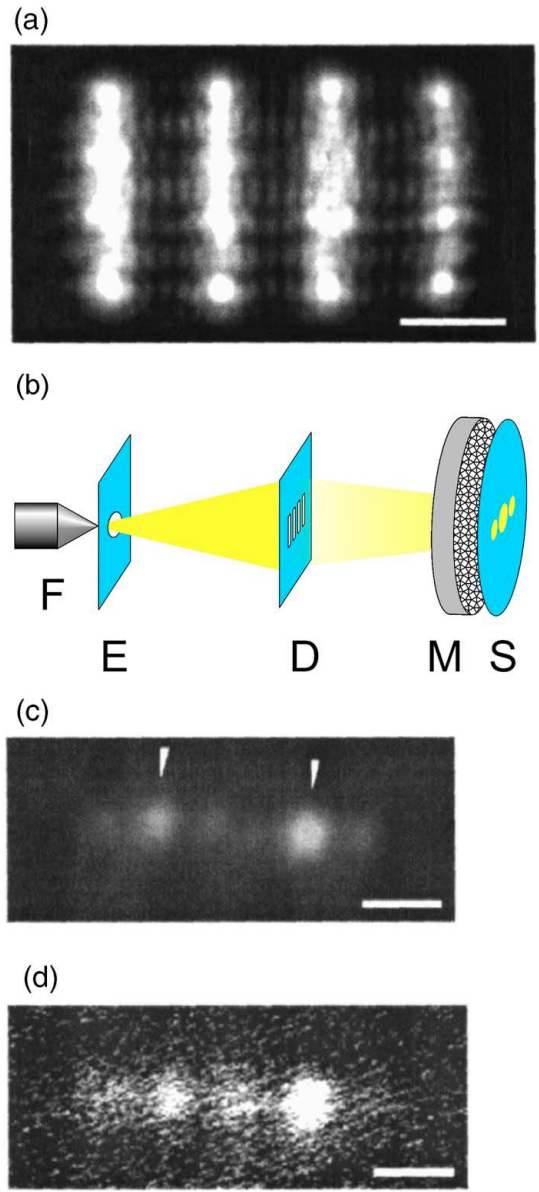

Figure 2. Electron diffraction experiments. (a) A LEEPS image of a diffraction grating is shown here, taken with 69 eV electrons. Patterns other than the shadow images of the actual holes are due to Fresnel diffraction. Scale bar: 100 nm. (b) The experimental setup consists of a field emitter (F) followed by an extractor (E), the diffractive element (D), a microchannel plate (M), and a phosphor screen (S). The electron beam was generated from a W(111) single crystal field emitter. A carbon thin film extractor with a 1.2 μm diameter hole (Quantifoil R1.2/1.3) was located 4 μm away from the emitter. The diffraction grating was located 360 μm away from the electron source. Because of the small diffraction angle, the image on the phosphor screen was magnified by an optical microscope, and subsequently recorded by a charge-coupled device (CCD) camera. (c) A diffraction pattern produced by two diffraction gratings. Diffraction patterns of each of the gratings exhibits zeroth order spots (indicated by arrows) in between two first order diffraction spots. The electron energy was 149 eV. Scale bar: 3 mrad. (d) When the electron energy was lowered to 90 eV, the two first order diffraction spots merged. Scale bar: 3 mrad.

We are interested in how well the above scheme would work in practice. We address this question by numerical simulations combined with Bayesian statistical analysis. Instead of demonstrating improved structural information retrieval, here we choose a simple example that clearly illustrates the essence of the method. Hence, we set a moderate goal of recognizing the orientation of 70S ribosome (6) out of two possible orientations "right" and "wrong," differing by 90°. Note that determination of the orientation of a molecule is an important issue also for the conventional structural determination methods. Our hypothesis is that the ribosome is in the right orientation, and we compute how many electrons are needed to verify the hypothesis. Two electron energies of 100 and 15 keV were considered as shown in Figs. 3(b) and 3(c), respectively. The first case represents a typical electron energy widely used in biological transmission electron microscopy. At this energy 70S ribosome can be modelled as a pure phase object. The second situation is less conventional.

However, it turns out that our method works better with 15 keV electrons, where 70S ribosome behaves as both a phase and amplitude object. The setup of the simulated experiment is shown in Fig. 3(a). The geometrical configuration chosen here was such that the structure of the diffractive element is compatible with current FIB technology. The structures of the diffractive elements are either fully transparent or opaque, with pixel sizes of 20 and 40 nm for 100 and 15 keV electrons, respectively. The random electrons arrival positions, obeying the probability distribution given by the intensity on the screen, were generated by a Monte Carlo method. In both energy cases, when the object is in the right orientation, after a significant amount of electrons has been detected, a spot can be clearly seen on the screen and no special statistical analysis is required. However, in practice the fragile molecules would be destroyed before the spot can be formed. To obtain high confidence of the object recognition with a few electrons only, Bayesian statistical analysis was applied to each detected electron. Our results, based on 500 numerical experiments, show that 4±3 electrons are needed at both, 15 and 100 keV, to identify the right object with 95% confidence, as shown in Figs. 3(b) and 3(c). However, the absorption by the ribosome and the diffractive element must be taken into account, which changes the number of required electrons to 51±38 electrons at 15 and 607±438 electrons at 100 keV. Nevertheless, there is a good prospect for reducing the dose down to a few electrons by placing the diffractive element in between a convergent wave electron source and the specimen. In this case, use of an off-axis configuration would further eliminate irradiation of the specimen by unnecessary electron waves that are due to the zeroth order diffraction.

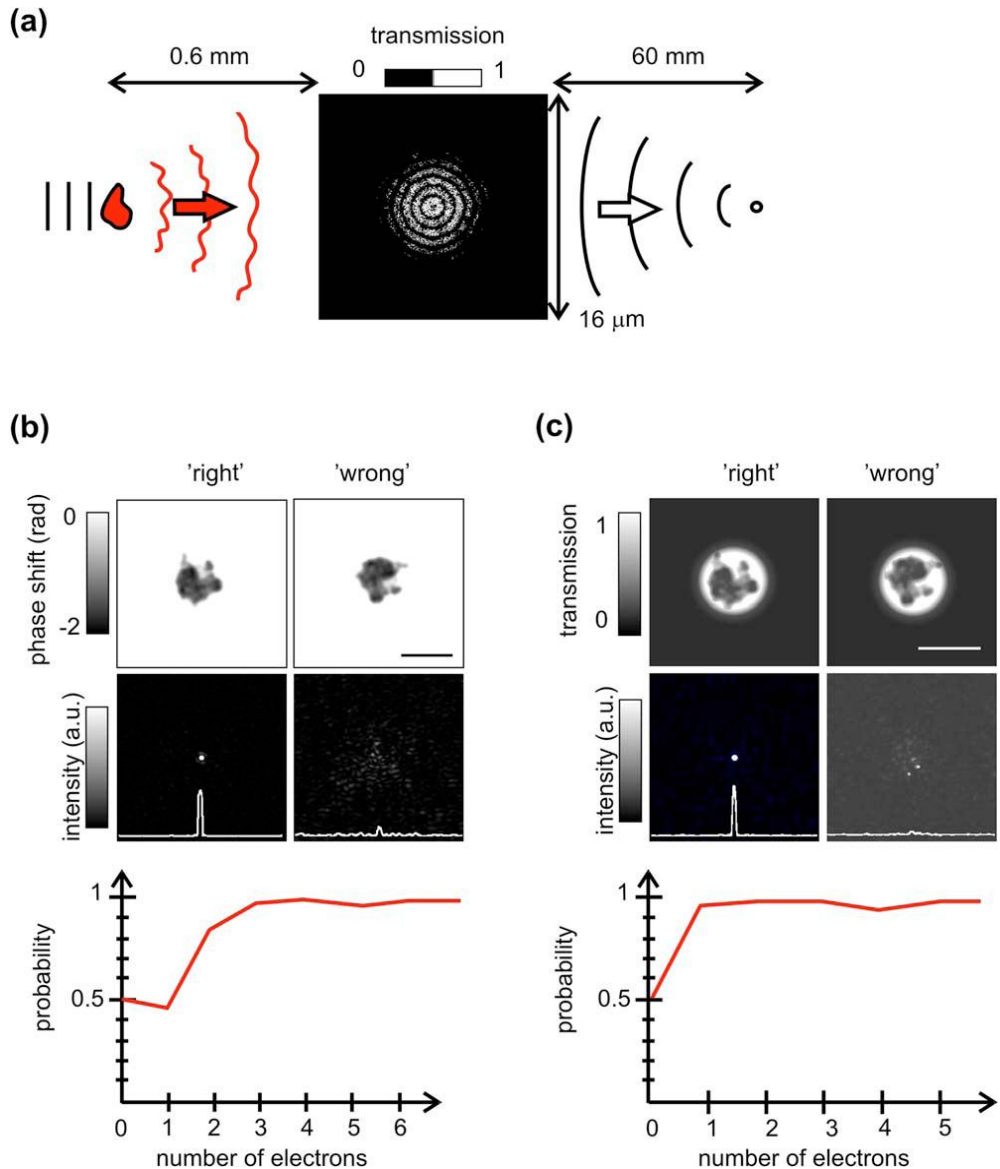

Figure. 3. Numerical experiments for recognizing different orientations of a ribosome S70 molecule. (a) An illustration of the experimental setup used in the simulations. The molecule is assumed to be in a vacuum. In practice, this situation could be approximated by freeze-drying method (7). The incident electron beam diameter is slightly larger than the molecule. The diffractive element transmission map corresponds to the simulated diffractive element for 15 keV electrons. Panels (b) and (c) represent results of numerical experiments for 100 and 15 keV electrons, respectively. The upper pictures show the transmission or phase shifts of the specimens (8). Scale bar: 20 nm. The pictures in the middle row show the intensity distribution on the screen. The curves in these pictures are the profiles of the intensity in the middle of the screen. A single spot is observed on the screen when the object is in the "right" orientation (left pictures) and a spread distribution is visible when the ribosome is rotated by 90° (right pictures). The graphs in the lower row show the confidence level for the hypothesis about the molecular orientation as a function of the number of detected electrons in a typical numerical experiment.

## Acknowledgement

The authors thank Dr. Gregory B. Stevens for discussions and Cornel Andreoli for experimental assistance.